\documentclass[%
reprint,
amsmath,amssymb,
aps
]{revtex4-1}

\usepackage{graphicx}
\usepackage{dcolumn}
\usepackage{bm}
\usepackage{wasysym}
\usepackage{comment}
\usepackage{pgfplots}

\begin{document}

\title{Interplay between optical, viscous and elastic forces on an optically trapped Brownian particle immersed in a viscoelastic fluid.}

\author{P. Dom\'inguez-Garc\'ia}
 \affiliation{Dep. F\'isica Interdisciplinar, Universidad Nacional de Educaci\'on a Distancia (UNED), Senda del Rey 9, Madrid 28040, Spain} 

\author{L\'{a}szl\'{o}  Forr\'{o}} 
\affiliation{Laboratory of Physics of Complex Matter, Ecole Polytechnique F\'{e}d\'{e}rale de Lausanne (EPFL), 1015 Lausanne, Switzerland}

\author{Sylvia Jeney}
\affiliation{Laboratory of Physics of Complex Matter, Ecole Polytechnique F\'{e}d\'{e}rale de Lausanne (EPFL), 1015 Lausanne, Switzerland}

\begin{abstract} 
We provide a detailed study of the interplay between the different interactions which appear in the Brownian motion of a micronsized sphere 
immersed in a viscoelastic fluid measured with optical trapping interferometry. To explore a wide range of viscous, elastic and optical forces,
we analyze two different viscoelastic solutions at various concentrations, which 
provide a dynamic polymeric structure surrounding the Brownian sphere. 
Our experiments show that, depending of the fluid, optical forces, even if small, slightly modify the complex modulus at low frequencies.  
Based on our findings, we propose an alternative methodology 
to calibrate this kind of experimental set-up when non-Newtonian fluids are used. 
Understanding the influence of the 
optical potential is essential for a correct interpretation of the 
mechanical properties obtained by optically-trapped probe-based studies of biomaterials and living matter.
\end{abstract}

\maketitle

Optical tweezers \cite{ashkin_applications_1980} have been widely 
used during the last decades for the study of forces and mechanical properties in micro-nano
environments, with special emphasis 
on living cells \cite{Ashkin1990, Tolic2004} and in combination with microrheology \cite{mason_optical_1995, mason_particle_1997, Waigh2016}, 
allowing the measurement of the mechanical
properties of complex fluids by tracking the movement of an optically-trapped micro-nano spherical probe bead. However, some doubts have
recently arisen \cite{Tassieri2015} about the viability of microrheological studies in living cells 
when using optical tweezers.
The gel's low-frequency elastic 
modulus should be greater than the elastic behavior from the optical trap, i.e., $G'(0)>G'_k = k/6\pi a$, where
$k$ is the spring constant from the optical trap modeled as a restoring force. 
In that case, the gel's elastic force traps the bead instead
of the optical tweezers. 
Therefore, the interplay between elastic forces, 
both external and from the fluid, 
is essential to understand the influence of the 
optical potential when the  
mechanical properties of biomaterials and living matter are estimated from the motion of optically-trapped Brownian probes.

The main objective of this work is to study the forces which act in the Brownian particle when it is immersed in known viscoelastic
fluids. 
Here, we study this problem experimentally, using optically trapped microbeads 
immersed in aqueous solutions of poly(ethylene oxide) or wormlike micelles at different concentrations. 
Both solutions have been well characterized in the past \cite{Berret1993, Cardinaux2002, Sprakel2007, DominguezGarcia2014} 
and display strong viscoelastic properties. We use these fluids to analyze the interplay between the applied
optical potential, considered harmonic in the central region of the potential well \cite{Svoboda2002, Richardson08},
the stochastic thermal force which generates Brownian motion \cite{Xin2016} 
and those produced by the fluid, which can be studied by its complex modulus through its elastic and viscous parts. 

We experimentally explore the motion of spherical microbeads with
radius $a = 0.94$ $\mu$m in an experimental optical trapping interferometry (OTI) set-up  
composed by an optical trap \cite{ashkin_applications_1980} ($\lambda = 1064$ nm) 
and an interferometric position detection system \cite{jeney_s_monitoring_2010},
allowing the measurement of bead position immersed in a fluid with nanometer accuracy. 
The beads are composed of plain, highly cross-linked
melamine resin with density $\rho = 1570$ kg/m$^3$ and the experiments are performed at $T= 294$ K.
The volume fraction of the particle in the used solution was $10^{-3}$ wt\%, 
which ensured a distance of at least 15 $\mu$m between the particle studied and its next neighbours. 
The absence of other particles close to the one studied was checked by microscopy and by moving the piezo-stage in all three dimensions.
The optical trap locates the bead in the middle of the sample chamber, avoiding the effects caused by
confinement \cite{faucheux_confined_1994, dominguez-garcia_p_single_2013} by exerting an external trapping force which
limits the bead dynamics \cite{lukic_motion_2007}. 
This experimental set-up reaches the short-time scale \cite{li_t_brownian_2013} at sampling rates in the MHz scale
and, therefore, high-frequency microrheology \cite{addas_microrheology_2004, DominguezGarcia2014}. 
The measurements are performed during $T=100$ s, providing
$N = 10^8$ points per experiment. The probe beads are immersed in
water, aqueous solutions of poly(ethylene oxide) (PEO) or wormlike micelle solutions \cite{CatesCandau90}.
These last ones are formed by surfactant cetylpyridinium
chloride (CPy$^{+}$ Cl$^{-}$) and sodium salicylate (Na$^{+}$ Sal$^{-}$).\footnote{The studied worm-like micelles solutions were prepared 
using published protocol\cite{Berret1993,Atakhorrami2006}. They had a molar ratio of [NaSal] = [CpyCl] = 100 mM.}
We use concentrations at the range $1$-$4$ wt\% for the micelle solutions and
of $0.5$-$15$ mg/ml for PEO solutions using the molecular weights $M_w = 495$ and $745$ kDa. 
The steady-state viscosities of the solutions, $\eta_0$, are
measured using a Rheometer MCR502 (Anton Paar, Austria) at $T = 294$ K.

\begin{figure}
\includegraphics[scale=1]{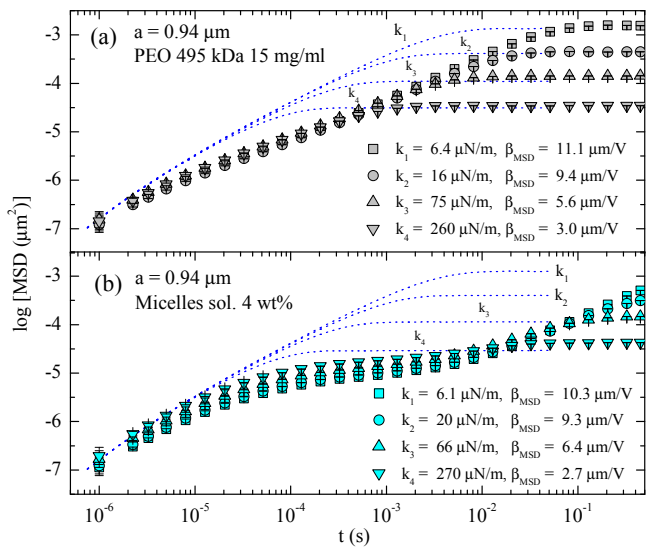}
\caption{\label{fig1}(Color online) Log-log representations of the 
1D MSDs for water (dotted lines) and viscoelastic fluids (points) 
using the four optical strengths available, where $k_1$ $<$ $k_2$ $<$ $k_3$ $<$ $k_4$. 
(a) PEO solution $M_w = 495$ kDa at 15 mg/ml (b) Wormlike micelle solutions at 4 wt\%. 
All data have been blocked in 5 points per decade.
}
\end{figure}

An important technical issue is calibration \cite{Capitanio2002}, i.e., the conversion from the
detected electrical signal, $V$, to the position of the bead in meter-units, $x$, 
defined by $\beta$, which is the position detector 
conversion factor $x=\beta V$. 
The calibration of the optical tweezers is considered complete
when the optical stiffness, $k$, is also obtained. 
Here, the calibration of the bead position $x$ is made using 
the methodology of the double flow chamber \cite{DominguezGarcia2014}.  
The cell is divided into two chambers: one contains water 
and the other encloses the viscoelastic fluid. 
In both fluids, we add a very small quantity of water with microbeads at a very low concentration. 
First, a microbead is captured with the trap in the chamber with water. The parameters $\beta$ and $k$ are calculated using 
Grimm's \textit{et al.} calibration method \cite{grimm_high-resolution_2012} for Newtonian fluids based on the hydrodynamic memory
appearing at low temporal scales. 
After the water measurement and using the same experimental conditions, a bead is trapped in the second chamber.
Hence, it is assumed that the calibration
factor, $\beta$, calculated using the signal from the first chamber, can be applicable to the data obtained from the second. 
An example of this calibration methodology when using viscoelastic fluids can be seen in fig. \ref{fig1}, where we plot
the measured mean-square displacements, defined as MSD$(t) \equiv \left<[x(t)-x(0)]^2\right>$, 
of microbeads with radius $a = 0.94$ $\mu$m 
for PEO and micelles aqueous solutions, using all the available laser powers.
The corresponding water measurements are also shown as dotted lines. 
The calibration parameters $\beta_{\text{MSD}}$ displayed in the legend in fig. \ref{fig1} 
are applied to the conversion from volts to meters in the viscoelastic fluids measurements. 

The MSDs from water show typical diffusive behavior in an intermediate temporal scale situated between the main influence of the optical trap 
and the prevalence of hydrodynamic effects  
at short time scales \cite{franosch_resonances_2011}, when $t < 10^{-5} \;\mu$s, 
where the solvent is predominant and the MSDs from the viscoelastic fluids collapse with the water measurements. 
Between these two temporal scales, the MSDs in water behave like MSD $\sim t^{\alpha}$ with $\alpha = 1$, 
while viscoelastic fluids show complex subdiffusive behavior with $\alpha < 1$
because of their viscoelastic behavior in that temporal range \cite{Hofling2013}. 
The behavior of PEO and micelle solutions is clearly different: 
Whereas PEO solutions show an almost constant $\alpha$ value related to a Rouse-Zimm behavior \cite{larson_structure_1999}, 
the micelle solutions have a variable $\alpha$ due to the dynamical polymeric behavior. 
At higher times, the optical trap is predominant at a characteristic time $\tau_k \equiv \gamma/k$, where $\gamma = 6\pi\eta a$ is the friction coefficient. 
There, the MSDs reach a plateau value which is equal to $\textrm{MSD}(\infty)=2k_B T/k$, where $k_B$ is the Boltzmann's constant, following the equipartition
theorem. 

\begin{figure}[t]
\includegraphics[scale=1]{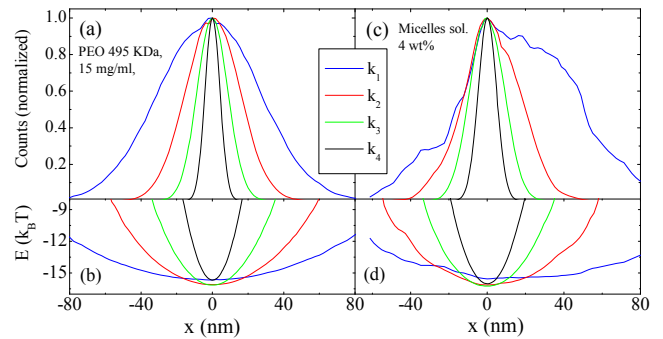}
\caption{\label{fig2}
(Color online) Gaussian distributions of the position $x$ (a) and (c) and curves of the optical trap potentials 
(b,d) for the data of fig. \ref{fig1}. 
 (a) and (b) PEO $M_w = 495$ kDa with $c = 15$ mg/ml, (c) and (d) micelle solutions at 4 wt\%.
Bin sizes are fixed to $0.2$ mV ($2.2$ nm for $k_1$). 
Distributions have been centered to $x=0$ and normalized to their respective maximum value for a better comparison between them.
}
\end{figure}

Regarding the influence of the optical trap on the bead's movement in different viscoelastic media, 
we analyze the optical force 
potential through Boltzmann statistics on the position of the Brownian particle.
If we obtain the probability density $p(x)$ from a experimental histogram of particle positions, 
the potential can be deduced as $E(x) = -k_B \,T\,\ln p(x) + C$,
where $C$ is a constant related to the potential offset and can be neglected \cite{Florin1998}.
In the case of a bead in an external optical trap, $p(x)$ will be a Gaussian distribution and $E(x)$ a quadratic polynomial. 
From this expression, we can obtain an experimental value of the optical stiffness, which we named $k_E$. 
An equivalent and straightforward calculation can be done by 
the variance of the Gaussian distribution: through the equipartition theorem 
$k_{\sigma^2} \equiv k_B\,T/\sigma^2$ where $\sigma^2 = \left< x^2 \right>$ is the variance of the probability density.

Fig. \ref{fig2} shows the distributions of the bead's positions and the corresponding trapping potentials 
for all the four trap stiffnesses available, using the same fluids as in fig. \ref{fig1}. 
The evolution of the position distributions indicate that, as the optical trap gets weaker, 
the influence of the viscoelastic fluid network surrounding the bead gets proportionally more important. 
Nevertheless, for most of the data, we obtain good Gaussian distributions and convincing harmonic quadratic 
curves for the trapping potential \cite{Richardson08}. 
Deviations from the Gaussian profiles translating even into asymmetric bead position distributions 
appear at lower optical forces, especially $k_1$, and more strongly in micelle solutions of higher concentrations than in PEO solutions.

In fact, for soft traps, the bead and the surrounding viscoelastic network are less confined by the trap and have more freedom to diffuse. 
The network around the bead may loosen up more easily and the bead can even ``jump'' in one direction through the structures forming the network. 
Such non-random ``jumping'' makes the histograms appear asymmetric. 
As the micellar network is per se more dynamic and hence less tight than the PEO mesh surrounding the bead, 
such ``jumps'' are more likely to appear in micellar solutions at comparable trap stiffness. 
Eventually, if measurements could last long enough the bead would, on average, jump in all directions 
through the network structures and the histogram would become symmetric and defined by the boundaries of the even softest harmonic trap. 
These longer times to explore the trapping potential boundaries are also needed because of the 
higher viscosity of the chosen worm-like micelle solution (table \ref{table:k}), 
which lowers the bead mobility.

Next, we compare the values of the optical stiffness, $k_1$, obtained both from the mean-square displacements calibration and from 
the experimental Gaussian distributions. 
The spring constants obtained from Boltzmann statistics are $k_{E}$ and $k_{\sigma^2}$,  
and the ones from the MSDs are $k_\text{MSD}$, assuming that all viscoelastic
data have been calibrated using the double chamber method. 
Table \ref{table:k} summarizes some of the calculations using PEO and micelle solutions with $k_1$ and bead size $a = 0.94$ $\mu$m. 
The different $k_1$ values are approximately similar, but they match specially well in the case of PEO with $M_w = 747$ kDa at $c = 15$ mg/ml.

\begin{table}[t]
\begin{ruledtabular}
\begin{tabular}{cccccc}
Fluid &$c$&$\eta_0$ &$k_\text{MSD}$  & $k_{E}$  & $k_{\sigma^2}$\\
\colrule
PEO 495 &$5$ &$3.1$ &$6.4$ & $5.44$ & $5.20$ \\ 
PEO 495 &$10$ &$6.7$ &$6.8$  & $6.0$ & $7.06$ \\
PEO 495 &$15$ &$11.9$ &$6.4$  & $5.5$ & $5.01$ \\
PEO 747 &$15$ &$68.2$ & $7.5$  & $7.9$ & $7.62$ \\
Micelles &2 &74 &$8.6$ &$5.1$ &$5.4$\\
Micelles &4 &380 &$6.1$ &$5.8$ &$4.3$\\
\end{tabular}
\end{ruledtabular}
\caption{\label{table:k} Trap strengths, $k$ ($\mu$N/m), calculated using the lowest
trap strength, $k_1$. PEO 495 and 747 are poly(ethylene oxide) water suspensions with $M_w = 495$ and $747$ kDa, respectively. 
Concentration values $c$
are given in mg/ml for PEO or wt\% for micelle solutions. Zero-shear viscosities $\eta_0$ are in mPa.s}
\end{table}

\begin{figure}[t]
\includegraphics[scale=1]{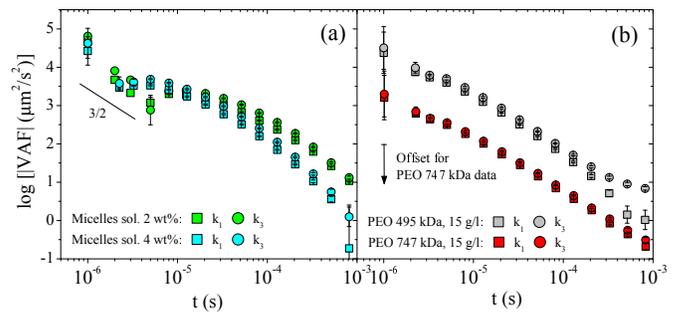}
\caption{\label{fig3}
(Color online) Log-log plots of $\left|\textrm{VAF}\right|(t)$
for different fluids and trap stiffnesses $k_1$ and $k_3$. 
(a) Wormlike micelle solutions at 2 and 4 wt\%. 
(b) PEO solutions at 15 mg/ml with $M_w = 495$ and $747$ kDa. 
An offset has been added to $747$ kDa data to better visualization.
All data have been blocked in 5 points per decade.
}
\end{figure}

Our experimental results show a clear different behavior between micelles and PEO solutions regarding the influence of the optical trap in the 
viscoelastic properties of these fluids. To check how the polymeric constituents of these fluids behave, we explore the absolute values of 
the velocity autocorrelation function, $\textrm{VAF}(t) \equiv \left<v(t)v(0)\right>$. 
If the VAF decay is not exponential, it indicates the presence of a network in the fluid surrounding the probe. 
Fig. \ref{fig3} shows $\left|\textrm{VAF}\right|(t)$ for micelles and PEO solutions at two trap stiffnesses. 
Both fluids follow an exponential decay, but they behave differently: micelle solutions change with concentration and optical trapping,
while PEO solutions are quite uniform for various molecular weights and optical forces. 
This observation indicates that the polymer networks are quite
different for each viscoelastic fluid. 
In these high-concentrated PEO solutions, 
the polymeric structures caged the bead and the optical trapping is not sensed by the particle, while in micelles the changing behavior of these
so-called living-like structures seem to be influenced by the optical forces.

\begin{figure}[b]
\includegraphics[scale=1]{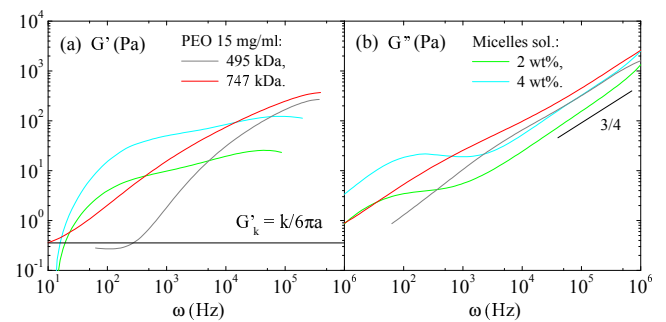}
\caption{(Color online) Microrheological calculations for aqueous solutions of PEO and micelles
using the lowest optical stiffness available, $k_1 \approx 6$ $\mu$N/m. (a) Storage modulus, $G'(\omega)$, (b) loss modulus $G''(\omega)$. 
\label{fig5}}
\end{figure}

As just seen, VAF$(t)$ is an excellent indicator of the short-time interaction between 
the probing microbead and the fluid \cite{grimm_high-resolution_2012,franosch_resonances_2011}, 
which, when viscoelastic, encages the probe. 
Depending on the caging strength of the fluid, softer optical traps have a smaller influence on this interaction 
at short times and VAF$(t)$ has a shape characteristic of the fluid. 
To investigate the elastic and viscous properties of the fluid further, we analyse the complex modulus at the softest trap.
Here, we apply the Mason-Weitz (MW) approach based on 
the Generalized Stokes-Einstein relation (GSER) \cite{mason_particle_1997}, 
concurrent with Mason's approximation \cite{mason_estimating_2000},
to obtain the complex modulus $G^*(\omega)$ from the measured MSDs. 
We correct these calculations to eliminate the inertia effects which appear in high-frequency microrheology \cite{DominguezGarcia2014}. 
We assume here that the optical force will appear in the microrheological results as a constant component in the elastic modulus, i.e.,
$G'_k = k/6\pi a$. Fig. \ref{fig5} shows
the results obtained for the same data as in figs. \ref{fig3} 
but using the lowest optical strength $k_1$.
In the low-frequency regime for the elastic modulus $G'(\omega)$, all values collapse to a single value of $G'(0) \sim 0.35$ Pa, 
corresponding to $k = 6\pi a\, G'(0) \sim 6$ $\mu$N/m, which
agrees with the measured value of $k_1$ by different methods. 
This means that the optical force is interacting with the fluids, 
which should affect the measurements of $k$ by Boltzmann statistics. 
The PEO solution with 747 kDa and 15 mg/ml shows the slightest deviation in its elastic
modulus, and only at very low frequencies it reaches $G'(0)$. 
The loss modulus, $G''(\omega)$, shows deviations
at low frequencies for micelle solutions, as expected from their changing polymeric behavior. 
At high frequencies, all data follow typical power-law
behaviors in $G''(\omega)$.

Our results show how PEO and micelle solutions are different viscoelastic fluids in their molecular structure. 
Their polymeric behavior marks the influence of the external optical forces in the studied fluids. PEO
solutions studied here are over the overlap concentration and as such, 
they are semi-dilute solutions forming a transient mesh. 
The correlation length or mesh size for PEO solutions with $M_w = 747$ kDa and $c = 15$ mg/ml is
$\xi = 43$ nm, similar to the mesh size of micelle solutions at 4 wt\%, which is $\xi = 42$ nm \cite{DominguezGarcia2014}. 
This mesh size is an order of magnitude smaller than the beam waist ($\sim 1$ $\mu$m) of the trapping laser.
The differences between the two fluids are then related to
the basic polymeric behavior. 
Micelle solutions are formed by very flexible chains which are in 
permanent process of breaking and recombination \cite{Berret1993}. 
The dynamical behavior of these surfactant systems has to be very influenced by 
the optical forces, even the smallest, specially when the polymer concentration is high enough. 
Then, the behavior of the storage modulus at
low frequencies for micelle solutions observed in fig. \ref{fig5} (a) is probably related to dynamical processes 
in the polymers structure because of the influence of the external optical force.

\begin{table}[t]
\begin{ruledtabular}
\begin{tabular}{ccccccc}
Fluid &$c$ &$a$  &$\eta_0$ &$\beta_\text{MSD}$  & $\beta_E$ & $\beta_{\sigma^2}$  \\ 
\colrule
PEO 495 &$15$ &$0.94$ &$11.9$ &$11.1$ & $10.7 \pm 0.9$ & $10.1 \pm 0.9$ \\ 
PEO 747 &$15$ &$0.94$ &$68.2$ &$10.2$ & $11.7 \pm 1.0$ & $11.5 \pm 1.0$ \\ 
Micelles &2 &$0.94$ &74 &$10.5$ & $10.3 \pm 0.9$ & $9.9 \pm 0.8$ \\
Micelles &4 &$0.94$ &380 &$10.2$ &$10.0 \pm 0.8$ & $8.7 \pm 0.8$ \\
\end{tabular}
\end{ruledtabular}
\caption{\label{table:beta} $\beta$ factors ($\mu$m/V) with
$\left<k_\text{H$_2$O}\right> = 6 \pm 1$ $\mu$N/m. 
Concentration values $c$ are given in mg/ml for PEO solutions or wt\% for micelle solutions and
$\eta_0$ in mPa.s}
\end{table}

Our measurements show that 
(i) the viscoelastic fluids studied here verify the condition exposed by Tassieri \cite{Tassieri2015} 
regarding the use of microrheology with optical traps in gels, 
(ii) the harmonic potential is observed when sufficient statistics are reached, and
(iii) $k$ values are not very different between them in these viscoelastic fluids, 
although their match depends on the type of fluid and concentration of polymer. 
Following these conclusions and as a final application of our study, we suggest an alternative method 
for calibration of bead position in optical-tweezers interferometry using non-Newtonian fluids.  
The present method does not require active modulation of the optical trap \cite{Fischer2007} 
if measurements times are long enough to allow the probe to explore the boundaries of the trap. 
This methodology consists in calculating averaged values 
from the calibration of MSDs for independent water measurements, $\left<k_\text{H$_2$O}\right>$.
Using the weakest laser power with bead size $a=0.94$ $\mu$m, we obtain $\left<k_1(\text{H$_2$O})\right> = 6 \pm 1$ $\mu$N/m. 
\footnote{The error of such averages is $\sim 20$ \% and agrees 
with the deviations previously reported for measurements of $k$ in different chambers \cite{Capitanio2002}.}
Then, we compare the $k$ values calculated using the thermal noise statistics with these averaged $\left<k_\text{H$_2$O}\right>$ 
to deduce the new calibration factors, $\beta$. 
We check this methodology with our viscoelastic fluids by calculating 
$\beta_E$ from the harmonic potential curve and $\beta_{\sigma^2}$ from the Gaussian distributions. 
The results are summarized in Table \ref{table:beta}, with a good general agreement between $\beta$ values. 
This methodology to obtain $\beta$ is simpler than other methodologies
to calibrate viscoelastic fluids \cite{Fischer2007}, 
although it needs an estimation of $G'(\omega)$ to evaluate 
the influence of the optical trap in the elastic behavior at low frequencies.

To summarize, we have explored thermal noise statistics in optical-tweezers interferometry experiments when probe microbeads move inside
viscoelastic media. 
In viscoelastic fluids, there is an appreciable 
interplay between the elastic and loss properties of the fluid at low frequencies 
and the external optical forces. 
We observe that, in most fluids, the optical trap is modifying the elastic modulus behavior at low frequencies. 
If that happens, the optical stiffnesses of the trap, $k$, calculated using different approaches, vary slightly.  
After the observation and evaluation of this interplay, we propose a calibration method when using
non-Newtonian fluids in an interferometry set-up.
This experimental work highlights an important point to have into account when using optical tweezers for
measuring complex fluids, specially biofluids or living materials, because it is usually ignored that 
the applied force may affect the internal structure of the material.

See supplementary material for additional information about experimental data and methodology.

PDG wants to acknowledge J. C. G\'{o}mez-S\'{a}ez for her proofreading of this text, 
M. A. Rubio and M. Pancorbo for fruitful comments, and to 
MINECO for financial aid by project FIS2013-47350-C5-5-R.

\end{document}